\documentclass{ws-ijmpa}
\usepackage[super,compress]{cite}
\usepackage{dcolumn}% Align table columns on decimal point
\usepackage{bm}% bold math
\usepackage{xcolor}
\usepackage{graphicx}
\usepackage{adjustbox}
\usepackage[utf8]{inputenc} % allow utf-8 input
\usepackage[T1]{fontenc}    % use 8-bit T1 fonts
\usepackage{hyperref}       % hyperlinks
\usepackage{url}            % simple URL typesetting
\usepackage{booktabs}       % professional-quality tables% blackboard math symbols
\usepackage{nicefrac}       % compact symbols for 1/2, etc.
\usepackage{microtype}      % microtypography
\usepackage{mathtools}
\usepackage{commath}
\usepackage[sc,osf]{mathpazo}

\let\norm\undefined % <-- "Undefine" \norm
\DeclarePairedDelimiter\norm{\lVert}{\rVert}
\usepackage{float}
\usepackage{varioref}
\usepackage[sans]{dsfont}
\usepackage{tikz}

\usepackage[makeroom]{cancel}
\DeclareMathOperator{\sign}{sign}
\newcommand{\Mathematica}{\textit{Mathematica\textsuperscript{\resizebox{!}{0.8ex}{\textregistered}}}}
\usepackage{xspace}
\def\8{\infty}
\def\oh{\frac{1}{2}}

\def\tt{\frac{2}{3}}

\def\d{\partial}
\def\i{\imath\,}

\def\dal{\partial_{\alpha}}
\def\dbe{\partial_{\beta}}
\def\dga{\partial_{\gamma}}

\def\undertext#1{\vtop{\hbox{#1}\kern 1pt \hrule}}
\def\ra{\rightarrow}
\def\Ra{\Rightarrow}

\def\VEV#1{\left\langle #1\right\rangle}

\def\diag#1{\hbox{diag}\left(#1\right)}
\def\dd#1{\frac{d}{d#1}}

\def\pbyp#1#2{\frac{\partial#1}{\partial#2}}

\def\bea{\begin{eqnarray} & &}
\def\eea{\end{eqnarray}}

\let\oldexp\exp
\renewcommand{\exp}[1]{\oldexp\left(#1\right)}

\def\NS{Navier-Stokes}
\def\KH{Kelvin-Helmholtz}

\def \BT{Burgers-Townsend}
\def \BRE{Birkhoff-Rott equation}

\def\CVS{\textit{CVS}}

\def\val{v_{\alpha}}
\def\vbe{v_{\beta}}

\def\rbe{r_{\beta}}
\def\rga{r_{\gamma}}

\def\oal{\omega_{\alpha}}
\def\obe{\omega_{\beta}}

\def\Xint#1{\mathchoice
   {\XXint\displaystyle\textstyle{#1}}%
   {\XXint\textstyle\scriptstyle{#1}}%
   {\XXint\scriptstyle\scriptscriptstyle{#1}}%
   {\XXint\scriptscriptstyle\scriptscriptstyle{#1}}%
   \!\int}
\def\XXint#1#2#3{{\setbox0=\hbox{$#1{#2#3}{\int}$}
     \vcenter{\hbox{$#2#3$}}\kern-.5\wd0}}

\def\Pint{\Xint/}

\DeclareMathOperator{\erf}{erf}
 
\renewcommand{\Re}{\textbf{Re }}
\renewcommand{\Im}{\textbf{Im }}

\begin{document}

% \preprint{APS/123-QED}

\title{Confined Vortex Surface and Irreversibility.\\
1. Properties of Exact solution}
\author{Alexander Migdal}
% \email{sasha.migdal@gmail.com}
\address{Department of Physics, New York University \\
  726 Broadway, New York, NY 10003}%Lines \date{=day}% It is always =day, today,
\maketitle
\begin{abstract}
  We revise the steady vortex surface theory following the recent finding of asymmetric vortex sheets (AM,2021). These surfaces avoid the \KH{} instability by adjusting their discontinuity and shape. The vorticity collapses to the sheet only in an exceptional case considered long ago by Burgers and Townsend, where it decays as a Gaussian on both sides of the sheet. In generic asymmetric vortex sheets (Shariff,2021), vorticity leaks to one side or another, making such sheets inadequate for vortex sheet statistics and anomalous dissipation.
  We conjecture that the vorticity in a turbulent flow collapses on a special kind of surface (confined vortex surface, or 
  \CVS{}), satisfying some equations involving the tangent components of the local strain tensor.
 
  The most important qualitative observation is that the inequality needed for this solution's stability breaks the Euler dynamics' time reversibility. We interpret this as dynamic irreversibility. We have also represented the enstrophy as a surface integral, conserved in the Navier-Stokes equation in the turbulent limit, with vortex stretching and viscous diffusion terms exactly canceling each other on the \CVS{} surfaces.
  
  We have studied the \CVS{} equations for the cylindrical vortex surface for an arbitrary constant background strain with two different eigenvalues. This equation reduces to a particular version of the stationary \BRE{} for the 2D flow with an extra nonanalytic term. We study some general properties of this equation and reduce its solution to a fixed point of a map on a sphere, guaranteed to exist by the Brouwer theorem.
\end{abstract}

\section{Introduction}

The vortex surfaces recently came back to our attention after it was argued \cite{M20c, M21b} that they provide the basic fluctuating variables in turbulent statistics. 

Within the Euler-Lagrange equations, the shape $S$ of the vortex surface is arbitrary, as well as the density $\Gamma(\vec r \in S) $ parametrizing the velocity discontinuity $ \Delta \vec v = \vec \nabla \Gamma$. The corresponding vortex surface dynamics\cite{M88, AM89} represents a special case of the Hamiltonian dynamics in 2 dimensions with parametric invariance similar to the string theory. 

In conventional Euler-Lagrange dynamics, the vortex surface's shape is evolving, subject to \KH{} instability, while $\Gamma$ stays constant due to the Kelvin theorem.

The steady solution for the vortex surface $\mathcal S$, as we recently argued in  \cite{M20c, M21b} would correspond to a particular discontinuity density $\Gamma(\vec r) $ minimizing the fluid Hamiltonian.

This minimization is equivalent to the Neumann boundary condition for the potential flow $ \vec v(\vec r) = \vec \nabla \Phi_\pm(\vec r)$ inside and outside the vortex surface
\begin{eqnarray}
   \d_n \Phi_+(\mathcal S)= \d_n \Phi_-(\mathcal S) =0
\end{eqnarray}

The local normal displacement $z$ of the surface\footnote{the tangent displacement is equivalent to reparametrization of the surface and as such can be discounted \cite{M21a}.} satisfies the Lagrange equation
\begin{eqnarray}
   \d_t z = v_n  =\d_n v_n z + O(z^2)
\end{eqnarray}

The positive normal strain $S_{n n} = \d_n v_n$ would lead to the \KH{} instability, but in case
\begin{eqnarray}
   S_{n n}(\mathcal S) <0,
\end{eqnarray}
this stationary vortex surface would be stable.

In the conventional analysis of the \KH{} instability \cite{BAT00} , Chapter 7, the variation of the displacement in the tangent direction is taken into account, which leads to a more general equation
\begin{eqnarray}
   \d_t z = -a z -\hat b z
\end{eqnarray}
where $\hat b$ is a linear differential operator. The normal strain $-a z$ was ignored.

The equation is linear and can be solved in Fourier space where $\hat b z \Ra \omega(k) \tilde z(k)$
\begin{eqnarray}
  \tilde z(k) = c_1 \exp{-(a + \omega(k)) t }
\end{eqnarray}

This $\omega(k)$ has two eigenvalues $\pm k \Delta v_t$, of opposite sign, and the negative eigenvalue leads to the \KH{} instability.
This  instability happens at large enough wave-vectors  $k \sim \frac{a}{\Delta v_t}$, which invalidates the Euler vortex sheets.

However, the smearing of the vortex sheet in the boundary layer by viscosity stabilizes it under certain conditions, which we discuss below.

We know some exact solutions of the \NS{} equations where a boundary layer replaces the vortex sheet with the width $h \sim \sqrt{\frac{\nu}{-a}}$.

\section{The strained flow}

Solution of the planar \NS{} equation in the thin layer around the surface of tangent discontinuity \cite{M21b} showed that there is a possible match between an arbitrary vortex surface and the old \BT{} planar vortex sheet\cite{BURGERS1948, TW51}. 

These sheets had the Gaussian profile of tangent vorticity as a function of the normal coordinate $z$, replacing the Euler solution's delta function.

This match assumes that the width $h$ of the Gaussian profile is much smaller than the surface's curvature radius. In that case, the \BT{} solution for the velocity in the local tangent plane can fill the velocity gap, as it becomes $\sign{z}$ in the limit of the vanishing viscosity.

This perfect match was "almost" proven: one parameter was left undetermined, namely, the normal derivative of the normal velocity. The width $h$ was related to this parameter as
\begin{equation}
    h = \sqrt{\frac{\nu}{-S_{n n}}}
\end{equation}

This normal strain must be negative for the existence of the \BT{} solution. It was conjectured in \cite{M21b} that this $-S_{n n}$ was some positive constant, uniform around the surface, scaling with viscosity in the same way as the gap $\Delta \vec v$.

This unproven conjecture was a weak point of the whole theory, as the full match could have revealed that it was variable around the sheet. Then it could violate the requirement of negativity, thereby leaking vorticity. 

As we show in this paper, the conjecture is not correct in general, but we find the replacement. This replacement -- the Confined Vortex Surface, or \CVS{} -- dramatically simplifies the whole theory of vortex surfaces. The random surfaces, which were the hardest part of the theory, are replaced by deterministic ones.

On these \CVS{} surfaces, the normal strain is indeed, a negative constant in the exact solution we find in this paper.

\section{Vorticity Confinement}

Here is what happened. It was recently observed \cite{M21a} that in addition to the \BT{} sheet with the symmetric Gaussian profile of vorticity, there is an asymmetric solution, expressed in the Hermite function with the negative fractional index. This asymmetric solution decays as a Gaussian on one side but only as a power on the other side of the sheet.

In other words, vorticity leaks from that sheet, unlike the \BT{} sheet where it collapses to a thin layer.

Later, another important observation was made \cite{KS21}. The asymmetric sheet turned out to be the general solution of the \NS{} equation for the constant background strain
\begin{equation}\label{strain}
    S_{\alpha \beta} = \oh(\dal \vbe + \dbe \val)
\end{equation}

The eigenvalues of the strain add up to zero in virtue of incompressibility, so there are two independent parameters here. 
\begin{equation}
    \hat S= \diag{\lambda_1,\lambda_2,-(\lambda_1+\lambda_2)}
\end{equation}
We can always assume that the third eigenvalue corresponds to the eigenvector in the normal direction.

The asymmetric solution exists when 
\begin{eqnarray}
     \lambda_1+\lambda_2 >0
\end{eqnarray}
as required by \KH{} stability.

The special case considered in \cite{M21a} corresponds to $\lambda_1 = \lambda_2 >0$. 
The vorticity of the generic solution is proportional to Hermite function
\begin{eqnarray}
 &&\omega \propto \exp{-\frac{z^2}{2 h^2}} H_\mu\left(\frac{z}{h \sqrt 
2}\right);\\
 && \mu = -\frac{\lambda_2}{\lambda_1 + \lambda_2};\\
 && \omega(z\ra +\infty) \propto (z)^\mu \exp{-\frac{z^2}{2 h^2}};\\
 && \omega(z\ra -\infty) \propto (-z)^\mu 
\end{eqnarray}
There is also a mirror solution with $z \Ra -z$.

For every finite $\lambda_1,\lambda_2$ vorticity decays at least on one side as a negative fractional power $|z|^\mu; \mu < 0$, which makes it unacceptable for the vortex surface statistics. 

The \BT{} solution corresponds to the exceptional case $ \lambda_1 >0,  \lambda_2 =0$. The solution reads
\begin{subequations}\label{BurgeSht}
\begin{eqnarray}
  &&\vec v = \{ a x, b S_h(z), -a z\};\\
  && S^0_{\alpha\beta} = \diag{a,0,-a};\\
  &&\vec \omega =\{-b S_h'(z),0,0\} ;\\
  &&S_h'(z) =\frac{1}{h \sqrt{2 \pi} } \exp{- \frac{z^2}{2 h^2}};\\
  && S_h(z) =\erf \left( \frac{z}{h \sqrt 2}\right);\\
  && a = \frac{\nu}{h^2};\label{aheq}
\end{eqnarray}
\end{subequations}

For the reader's convenience, we verified this solution analytically in various coordinate systems in a \Mathematica notebook \cite{MB}. 
In this case, the vorticity becomes Gaussian, and the velocity gap becomes an error function. In the limit of $h\ra 0$ the vorticity reduces to $\delta(z)$, and velocity gap reduces to $\sign(z)$.

These solutions for various ratios $\mu$ of eigenvalues were investigated in \cite{KS21}. An interesting case is negative $\lambda_2$ (super-Townsend in \cite{KS21}). In that case $ \mu < -1$ so that power decay is even stronger that in case of positive $\lambda_2$. 

The vorticity, in that case, starts from the peak at $z =0$ then decays as a Gaussian to some level, after which the power terms take over. These power terms in the super-Townsend case are negative so that the vorticity approaches zero from the opposite side after reaching the minimum.

\begin{figure}
    \centering
    \includegraphics[width=0.75\textwidth]{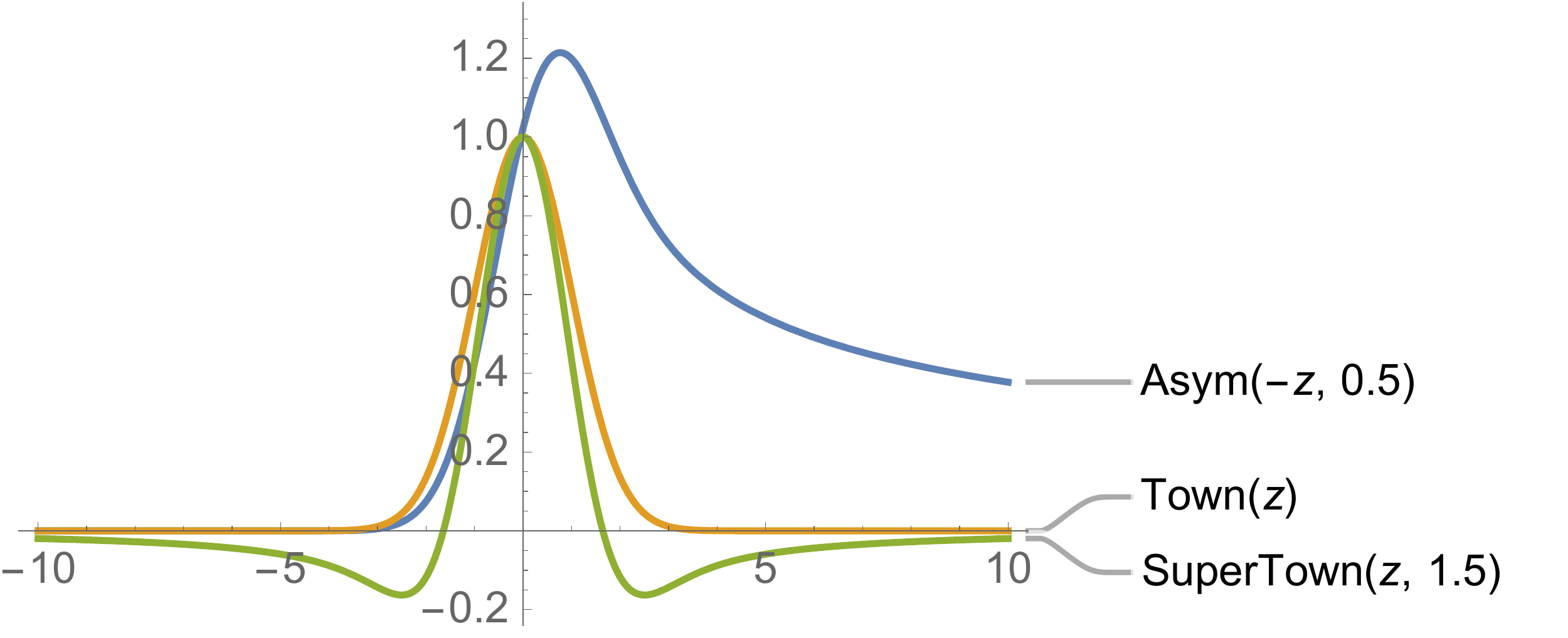}
    \caption{The vorticity profiles for asymmetric, Townsend and super-Townsend strains}
    \label{fig::VorticityProfiles}
\end{figure}

(Fig.\ref{fig::VorticityProfiles}).

The authors of \cite{KS21} also studied the time evolution. In the asymmetric case ($a <1$ in \cite{KS21}, or $\lambda_2 >0$ in our notation), the peak of vorticity decays. This decay happens due to the leaks on one or both sides.  The solution moves from the steady state to zero vorticity.

The super-Townsend case $a >1$ or $-\lambda_1 <\lambda_2 <0$  turns out to be unstable. Rather than decaying, the vorticity accumulates in the negative pockets and leaks from there. This solution moves away from the steady state, making it unstable.

Only the exceptional Townsend case $\lambda_2 =0$ proves to be stable (up to some finite-size effects). Vorticity in the Gaussian peak does not decay, nor it grows on the sides; it stays the same.

The vortex sheets with Gaussian core and small pockets outside, like in the super-Townsend case, were also observed in simulations of a real isotropic turbulence \cite{KS21}.  

Presumably, these observed pockets are finite Reynolds effects; they will disappear in the extreme turbulence when the peak grows to infinity.

If we would like to match the Euler vortex surface with the \NS{} equation in a boundary layer, we must seal the leak of vorticity, which requires one vanishing eigenvalue. Furthermore, we need the normal strain to be negative to squeeze the vorticity into the sheet.

In that case, the sorted eigenvalues  will be $(\lambda, 0, -\lambda)$. We can always choose the local coordinate system so that the normal to the surface is parallel to one of the strain's remaining eigenvectors. 

One of the two remaining eigenvectors (corresponding to the vanishing eigenvalue) must be directed along the velocity gap ( $y$- axis in the \BT{} solution), which provides another condition for stability.

Then we would have $S_{n n} =\pm\lambda$. The flow direction change while preserving its geometry would change this sign in front of $\lambda$. Only the negative sign is acceptable, as only then the real solution for the width $h$ would exist.  The velocity gap points towards the vanishing component of the strain in its diagonal frame.

Generalizing these observations, we suggest a new scenario for extreme turbulence, which we call Confined Vortex Surface, or \CVS{}.

\medskip
\fcolorbox{red}{yellow}{%
    \parbox{0.9\textwidth}{%
        The turbulent vorticity collapses along the closed surface where the normal velocity vanishes, the tangent strain annihilates the tangent velocity gap but
        the normal strain is negative.
        This required negative normal strain breaks the time reversal but preserves the space symmetries, including parity.
        }
}
\medskip

These conditions for the stability can be written in invariant form
\begin{eqnarray}\label{CVS}
    && \vec \sigma \cdot \vec v =0;\\
   && \hat S \cdot \Delta\vec  v =0;\\
   && \vec \sigma \cdot \hat S \cdot \vec \sigma < 0;
\end{eqnarray}
where $\vec \sigma $ is a local normal vector to the vortex surface.

In the rest of this paper, we study the \CVS{} and its implications for the vortex surface dynamics.

\section{The Basic Equations}

The steady closed vortex surface $\mathcal S$ can be treated within the framework of hydrostatics, as it was recently advocated in our paper \cite{M21a}.

In the 3D space inside and outside the surface $ \mathcal S^\pm: \d \mathcal S^\pm = \mathcal S$ there is no vorticity, so the flow can be described by a potential $\Phi_\pm(\vec r)$
\begin{eqnarray}
   \val(\vec r) = \dal  \Phi_\pm(\vec r) ; \; \forall \vec r \in S^\pm
\end{eqnarray}
The incompressibility $\dal \val =0$ would be satisfied provided both potentials satisfied Laplace equation with the Neumann boundary conditions at the surface. The stationary surface requires vanishing normal velocity \cite{M20c,M21b}
\begin{eqnarray}
   &&\dal \val(\vec r) = \dal^2 \Phi_\pm(\vec r) =0;; \; \forall \vec r \in \mathcal S^\pm\\
   && \d_n \Phi_+(\vec r) =  0;\; \forall \vec r \in \mathcal S\\
   && \d_n \Phi_-(\vec r) =0; \; \forall \vec r \in \mathcal S
\end{eqnarray}

 The potential is given by the Coulomb integral with dipole density $\Gamma$ plus a background constant strain potential
\begin{eqnarray}
   && \Phi_\pm(\vec r_0) = P(\vec r_0) -\frac{1}{4 \pi} \int_S  \Gamma(\vec r) d^2 \sigma_\alpha(\vec r) \dal \frac{1}{|\vec r - \vec r_0|};\\
   && d^2 \sigma_\alpha(\vec r) = e_{\alpha\beta\gamma} d \rbe \wedge d\rga;\\
  &&P(\vec r_0) = \sum_n -\frac{1}{4 \pi}\int_{\vec r \in S_n}  \Gamma_n(\vec r) d^2 \sigma_\alpha(\vec r) \dal \frac{1}{|\vec r - \vec r_0|}
 \end{eqnarray}

The $P$ term represents the contribution from similar dipole density terms from other remote vortex surfaces. 
Assuming these surfaces to be far away, we expand $P(\vec r_0)$ in Taylor series in $r_0$.
\begin{eqnarray}
P(\vec r_0)  \ra  \vec V_0 \cdot \vec r_0 + \oh \vec r_0 \cdot \hat W \cdot \vec r_0 + \dots
\end{eqnarray}
The leading term is some constant velocity $\vec V_0$ which can be eliminated by a Galilean transformation, so it does not have any physical effect.

We are going to stop the expansion at this first nontrivial $W$ term. Later we discuss the physical motivation for this truncation. This term represents the background strain created by other remote vortex structures, and as such, it slowly varies in space. 

The higher gradients of this background potential will be inversely proportional to the distance to these remote structures, which is the next order in the (assumed) small parameter $R_0/R$, where $R_0$ is the size of this vortex surface, and $R$ is the typical distance between such surfaces.

The  potential  satisfies the Laplace equation in $\mathcal S^\pm$ with continuous normal derivative of the potential at the surface.

The velocity field is computed as a gradient of this potential. Simple computations lead to the following expression
\begin{eqnarray}\label{Vint}
    &&\vec v(\vec r_0) = \hat W \cdot \vec r_0  + \frac{1}{4 \pi}\int d \Gamma \wedge d \vec r \times \vec \nabla \frac{1}{|\vec r - \vec r_0|}
\end{eqnarray}

The normal component of this velocity is continuous at the surface, as it can be checked by direct computation.

In the \CVS{} theory, we require that this normal velocity vanishes at every point on the surface
\begin{eqnarray}\label{Qeq}
 \vec  \sigma(\vec r) \cdot  \vec v(\vec r) =0 \forall \vec r \in \mathcal S
\end{eqnarray}

This condition is needed for a stationary surface in vortex sheet dynamics.

The gap in potential at the surface equals $\Gamma$ by construction. 
The velocity $\vec v = \vec \nabla \Phi$ has tangent discontinuity :
\begin{eqnarray}
    && \Delta \Phi = \Gamma;\forall  \vec r \in S;\\
    &&\Delta \vec v = \vec \nabla \Gamma(\vec r); \forall  \vec r \in S;\\
    && \vec \sigma \cdot \vec \Delta \vec v =0; \forall  \vec r \in S;
\end{eqnarray}

Taking the curl of velocity in the linear vicinity of a surface point in the tangent frame $x,y,z$, with $z$ directed at the local normal, we have
\begin{eqnarray}
    && \Delta v_i = \d_i \Gamma; i =1,2\\
    && \Delta v_z =0;\\
   && \omega_i(x,y,z) = \delta(z)  e_{i j }\d_i\Gamma ;\\
   && \omega_z(x,y,z) =0
\end{eqnarray}

It is straightforward to verify the divergence equation $\vec \nabla \cdot \vec \omega =0$. 

The following invariant integral generalizes this formula for an arbitrary continuous surface
\begin{eqnarray}
   &&\vec \omega(\vec r) = \int_\Sigma d\vec \Omega \delta^3\left(\vec X-\vec r\right);\\
   && d \vec \Omega \equiv d \Gamma\wedge d\vec X = d\xi_1 d \xi_2 e_{a b} \pbyp{\Gamma}{\xi_a} \pbyp{\vec X}{\xi_b} ;
\end{eqnarray}
The integration over local coordinates $x,y$ removes two of the three delta functions, which brings us to the above formula for the local vorticity in vicinity of the surface.

In Fourier space
\begin{eqnarray}
&& \vec \omega^F(\vec k) = \int d^3 r e^{\i \vec k \cdot \vec r} \vec \omega(\vec r) = \int_\Sigma d \vec \Omega e^{\i \vec k \cdot \vec X};\\
    &&\i \vec k \cdot \vec\omega^F(\vec k) =\int_\Sigma  d \Gamma \wedge d \vec X \cdot (\i \vec k) e^{\i \vec k \cdot \vec X}=
     \int_\Sigma d \Gamma\wedge d e^{\i \vec k \cdot \vec X} =\\
     &&\int_{\d \Sigma} d\Gamma e^{\i \vec k \cdot \vec X} =0;
\end{eqnarray}

From the point of view of the vortex sheet dynamics, the condition of vanishing normal velocity represents a linear integral equation for $\Gamma $ with symmetric kernel. 

It corresponds to the minimization of the quadratic Hamiltonian
\begin{eqnarray}
H_{0} = \int_{\vec r_1,\vec r_2 \in \mathcal S}  d \Gamma(\vec r_1) \wedge d \vec r_1  \cdot d \Gamma(\vec r_1) \wedge d \vec r_1 \frac{1}{8 \pi |\vec r_1 - \vec r_2| }
\end{eqnarray}
with extra linear term 
\begin{eqnarray}
H_{1} = \int_S \Gamma(\vec r)  d^2 \vec \sigma(\vec r) \cdot  \hat W \cdot \vec r
\end{eqnarray}

This equation is the first of the three basic equations of the \CVS{} theory.

The regular part of the strain, which is defined as the mean of the strains at two sides of the surface, is given  by
\begin{eqnarray}\label{SurfaceStrain}
&& S_{\alpha\beta}(\vec r) =\oh \dal \vbe(\vec r^+) + \oh \dal \vbe(\vec r^-)   + \left \{ \alpha \Longleftrightarrow \beta \right\};\;\forall \vec r \in  \mathcal S \text{;}
\end{eqnarray}
Here $\vec r^\pm$ corresponds to the limit taken from the outside/inside of the surface.

The vector \CVS{} equation
\begin{eqnarray}\label{GapEq}
   \hat S \cdot \vec \nabla \Gamma  =0
\end{eqnarray}
provides two more relations between the surface and the potential gap $\Gamma$.

Thus, we have three algebraic/differential equations for the unknown functions of coordinates: the parametric equation $\vec r = \vec X(\xi) =0$ of the surface $\mathcal S$ and the boundary data  $\Gamma(\xi)$ for the singular part of the potential.

At each point of the surface, the \CVS{} condition reduces to the following. One of the two eigenvalues of the tangent part of the strain tensor $S_{i j}, i,j =1,2$ must vanish. The velocity gap is directed along the eigenvector of this vanishing eigenvalue, i.e., the null vector of the surface strain.

Locally, the surface is described by two functions of two variables $ \Gamma(x,y), z(x,y)$, where $z(x,y)$ is the equation  for the normal coordinate as a function of the two tangent coordinates $x,y$. 

However, the \CVS{}{} equation is nonlocal, as it involves the derivatives of velocity, which are determined by the vorticity distribution over the whole surface. This leads to an integral equation for the unknown variables : $\Gamma, \mathcal S$.

This solution depends on the random constant strain tensor $W$, which comes from the "thermostat" of the remaining vortex structures in the turbulent flow. We discuss this issue in the next paper in this series. 

\section{Anomalous dissipation }
 
 The role of stable, steady solutions of the \NS{} equations is to provide the subspace of attractors in phase space. We expect the evolution of the solution to cover this subspace with a certain uniform measure like the Newton dynamics covering the surface of constant energy.
 
 The situation is more complex in turbulence. The energy $E$ of the fluid is not conserved, it is rather dissipated. This dissipation is proportional to the enstrophy
 \begin{eqnarray}
 \d_t E = -\mathcal E = -\nu \int d^3 r \oal^2 
 \end{eqnarray}
 
 We have a steady state of constant energy dissipation in the turbulent limit, with the constant energy supply from external forces on the system's boundary, like the submarine moving in the ocean or the water pumped into the pipe.
 
 Although ultimately related to the viscous effects, one can study this dissipation in the turbulent limit within the framework of the Euler-Lagrange dynamics.
 
 Large vorticity in certain regions of space where dissipation predominantly occurs will offset the vanishing factor $\nu$ in front of the enstrophy. 
 
 Our theory starts with the conjecture that dissipation occurs in vortex surfaces, like the \BT{} sheet. In that case, the dynamics are described by the Euler-Lagrange equations everywhere except the turbulent layer surrounding the vortex sheet, as we discussed above.

Using the \BT{} vortex sheet solution in the linear vicinity of the local tangent plane to the surface, as in \cite{M20c, M21b} we get the following integral for the dissipation
\begin{eqnarray}
\mathcal E  \ra \nu  \int_D d^2 \xi \sqrt{g} (\vec \nabla \Gamma)^2 \int_{-\infty}^{\infty} d \eta(\delta_h(\eta))^2;
\end{eqnarray}
  
Here $\eta$ is a local normal direction to the surface and
\begin{eqnarray}
  \delta_h(\eta) = \frac{1}{h \sqrt{2 \pi}}\exp{- \frac{\eta^2}{2 h^2}}
\end{eqnarray}
is the normal distribution.
The (local!) width $h$ is determined by
\begin{eqnarray}
  h = \sqrt{\frac{\nu}{ \lambda}}
\end{eqnarray}

Here $\lambda = - S_{n n} = S^i_i$. 

Naturally, we assume that the local normal $\vec \sigma$ points on the third axis of the strain, where the finite eigenvalue is negative.

An important new phenomenon is the possibility of the variation of this eigenvalue along the surface.

The square of the Gaussian is also a Gaussian:
\begin{eqnarray}
  &&\delta_h(\eta)^2 = \frac{1}{ 2h \sqrt{\pi} }\delta_{\tilde h}(\eta);\\
  && \tilde h = h/\sqrt{2}.
\end{eqnarray}
In the limit $\tilde h \ra 0$ we are left with the surface integral
\begin{eqnarray}\label{DissipationStrain}
  \mathcal E =  \frac{\sqrt{\nu}}{ 2\sqrt{\pi} } \int_D d^2 \xi \sqrt{g}\sqrt{-S_{n n} } (\vec \nabla \Gamma)^2;
\end{eqnarray}

In the previous paper \cite{M21b} we also found an expression for the time derivative of the viscosity anomaly for the Euler equation. 

Repeating these steps with our local width $h$ we find (fixing wrong sign in \cite{M21b})
\begin{eqnarray}
    &&(\d_t \mathcal{E})_{Euler} =  2\nu \int d^3 r \left(-\dbe \left(\oh \vbe \oal^2\right) + \oal  \obe  \dbe \val \right) \ra\nonumber\\
    && +\frac{\sqrt{\nu}}{ 2\sqrt{\pi} } \int_D d^2 \xi \sqrt{g}\sqrt{S^i_i}  \tilde \d^i \Gamma S_{i j}\tilde \d^j \Gamma ;\\
    &&  \tilde \d^i = \epsilon^i_k \d_k
\end{eqnarray}

We did not add the viscous term of the \NS{} equation in that paper, which was a mistake. Adding it now, we get, after a  simple algebra
\begin{eqnarray}
    &&(\d_t \mathcal{E})_{NS} = (\d_t \mathcal{E})_{Euler} + 2 \nu^2 \int d^3 r \; \vec \omega \cdot\vec \nabla^2 \vec \omega \ra \nonumber\\
    && +\frac{\sqrt{\nu}}{ 2\sqrt{\pi} } \int_D d^2 \xi \sqrt{g} \sqrt{S^k_k} \tilde \d^i \Gamma \tilde \d^j \Gamma \left(S_{i j}- S^k_k g_{i j}\right)
\end{eqnarray}
We transform this expression into a simpler one using the properties of two-dimensional tensors
\begin{eqnarray}
(\d_t \mathcal{E})_{NS} =-\frac{\sqrt{\nu}}{2 \sqrt{\pi} } \int_D d^2 \xi \sqrt{g} \sqrt{S^k_k}\d^i \Gamma S_{i j}\d^j \Gamma
\end{eqnarray}
In the general case, this expression is not zero and could have an arbitrary sign, because the tangent eigenvalues can be both signs, given that their sum is positive.

However, in the \CVS{} case, the Euler and \NS{} terms in time derivative of the enstrophy cancel so that the surface dissipation integral is conserved!

The \CVS{} equation, derived from the microscopic stability of the Euler dynamics in the turbulent limit, also provides a new turbulent motion integral: the surface dissipation \eqref{DissipationStrain}. 

Unlike the energy functional, this is an \textit{integral of motion of the \NS{} dynamics}, but not the Euler dynamics.

The \CVS{} surface is unique in having these two effects: vortex stretching and diffusion  cancel each other, making the surface enstrophy an integral on motion of the \NS{} dynamics in the turbulent limit.

Naturally, the enstrophy is constant on every steady solution of the \NS{} equation, simply by the definition of a steady solution.
What is not trivial and is quite fortunate for the vortex sheet dynamics is that the \textit{ surface} dissipation is conserved.

The energy could be dissipated all over the space, not just on the vortex surface. This phenomenon, indeed, happens for every vortex sheet except the \CVS{}. The time derivative of the enstrophy is either negative on these surfaces, which means that vorticity is leaking outside, or it is positive, meaning an instability of the surface.

The total dissipation integral, in that case, would not be dominated by a surface, which would make the theory incomplete.

The simple idea that the turbulent statistics is a Gibbs distribution with surface dissipation in the effective Hamiltonian was advocated in our recent paper \cite{M21b}. The conservation of the surface dissipation was unknown at that time, which was the problem.

Now we have this problem solved in \CVS{} dynamics.

Conservation of the surface integral for the energy dissipation puts this integral in the unique position of an effective Hamiltonian for turbulent  statistics, obeying the Liouville theorem of conservation of measure in phase space.
\begin{eqnarray}
\mathcal Z = \sum_{CVS} \exp{- \beta \frac{\sqrt{\nu}}{ 2\sqrt{\pi} } \int_D d^2 \xi \sqrt{g}\sqrt{-S_{n n} } (\vec \nabla \Gamma)^2} 
\end{eqnarray}
By the sum over \CVS{} we imply integration over the space of parameters of the \CVS{}  solutions.

Just like the Gibbs distribution for the Newtonian dynamics, this distribution represents the stable fixed point of the evolution of the Hopf functional for the \NS{} dynamics \cite{M20c}. The significant difference is that with the \CVS{} constraints, this fixed point is less than a field theory.

These constraints reduce the arbitrary surface and arbitrary potential gap $\Gamma$ to some family of solutions with a finite number of parameters. This family of solutions is no longer the functional phase space of the general vortex sheet dynamics \cite{M88,AM89}.

This finite-dimensional attractor could be a Holy Grail of the turbulence quest, but first, we must thoroughly investigate this hypothesis.

In this paper, we make the first step by finding a  family of exact solutions of the \CVS{} equations with cylindrical geometry, parametrized by two eigenvalues of a background strain tensor.
The analytic form of the solution is not known, but its existence is guaranteed by the Brouwer theorem for a sphere\cite{BrT}.

The turbulent statistics of these \CVS{} vortex sheets is the subject of the second part of this study, to be published soon.

\section{ Cylindrical Vortex Surface}

Let us go to the coordinate frame where the strain $\hat W$ is diagonal
\begin{eqnarray}
   \hat W = \diag{p + q, p - q, - 2 p};
\end{eqnarray}

Now, we consider the equation for the angular profile $r = R(\theta)$ of the surface in complex cylindrical coordinates.

\subsection{Complex Analysis}

Let us treat this problem from a complex analysis point of view. We know that in two dimensions $x,y$ the harmonic potential is represented by real part of a holomorphic function of $x + \i y$. Let us see how this happens for the double layer potential in cylindrical geometry.

The surface integral in \eqref{Vint} reduces to the following (with $\vec r_0  = (x,y)$ )
\begin{eqnarray}
&&e_{\alpha\beta\gamma}\int d \Gamma \wedge d \rbe  \dga \frac{1}{|\vec r - \vec r_0|} =\nonumber\\
&& e_{i j}\int d \Gamma(\theta) \int_{-\infty}^{\infty} d z  \frac{(C_j(\theta) - r_{0,j})}{\left(|\vec C(\theta) - \vec r_0|^2 + z^2\right)^{\frac{3}{2}}} =\nonumber\\
&&=  2 e_{i j}\int d \Gamma(\theta) \frac{(C_j(\theta) - r_{0,j})}{|\vec C(\theta) - \vec r_0|^2} \nonumber \\
&&= 2 e_{i j} \d_j \int d \Gamma(\theta) \log \left| \vec C(\theta) - \vec r_0 \right|
\end{eqnarray}

Introducing the complex velocity and complex loop, we get our holomorphic function of $ \eta = x + \i y$
\begin{eqnarray}
&& v_x - \i v_y = V(\eta) = \frac{1}{2 \pi \i}\int\frac{d \Gamma(\theta)}{\eta - C(\theta)};\\
&&C(\theta) = C_x(\theta) + \i C_y(\theta)
\end{eqnarray}
This complex velocity is a derivative of a complex potential
\begin{eqnarray}
   &&f(\eta) = \frac{1}{2 \pi \i}\int d \Gamma(\theta) \log \left(\eta - C(\theta) \right);\\
   && V(\eta) = f'(\eta);
\end{eqnarray}
By design, $f(\eta)$ must be holomorphic outside the profile loop $\mathcal C$.  This is achieved  by analytic continuation from the outside region of $\eta \in C^+$. When the point $\eta$ moves into the inner region $C^-$ it crosses the loop.

In the vicinity of the crossing point $\theta = \theta_0$ one can expand $C(\theta)\ra C(\theta_0) + C'(\theta) (\theta- \theta_0)$ and use the formula
\begin{eqnarray}
   \frac{1}{C'(\theta_0)\left(\theta-\theta_0 - \i \epsilon\right)}  = \frac{1}{C'(\theta_0)\left(\theta-\theta_0 + \i \epsilon\right)} + \frac{2 \i \pi}{C'(\theta_0)} \delta(\theta-\theta_0) 
\end{eqnarray}

This way we find the tangent gap  of the complex velocity
\begin{eqnarray}
   \Delta V(C(\theta_0)) = \frac{\Gamma'(\theta_0)}{C'(\theta_0)}
\end{eqnarray}
Note that we do not treat $C(\theta)$ as a boundary value of a holomorphic function. This requirement is not necessary for the \CVS{} equations, moreover, it is not compatible with these equations, where the real and imaginary parts of $C(\theta)$ enter as independent functions of $\theta$.

The normal vector in complex notation
\begin{eqnarray}
 \sigma =  \i C'(\theta)
\end{eqnarray}
The normal projection of velocity field
\begin{eqnarray}
   &&v_x \sigma_x +  v_y \sigma_y = \Re (v_x - \i v_y) \sigma =
   \frac{\Im \left( (v_x - \i v_y) C'(\theta) \right)}{|C'(\theta)|}
\end{eqnarray}
The above equation for normal velocity becomes
\begin{eqnarray}\label{Neumann}
   \Im\left( (p C^*(\theta) + q C(\theta) +V(C(\theta)) C'(\theta)\right) =0;\forall \eta =C(\xi)
\end{eqnarray}

The problem reduces to finding this  loop and the density $d \Gamma(\theta) = \Gamma'(\theta) d \theta$.

\subsection{ \CVS{} equations}

Let us consider the \CVS{} equation \eqref{GapEq}, which relates $f''$ to $f'$.

This equation requires the computation of the strain related to the complex potential $f(\eta)$. 
The strain in \eqref{SurfaceStrain} is a $3 X 3$ matrix 
\begin{eqnarray}
   \hat S =  \left(
\begin{array}{ccc}
 \Re\left( f''\right)+p+q & - \Im\left(f''\right) & 0 \\
 -\Im\left(f''\right) & -\Re\left(f''\right)+p-q & 0\\
  0                  &0                         &-2 p\\
\end{array}
\right)
\end{eqnarray}

The null vector equation $\hat S\cdot \Delta \vec v =0$ provides the following complex equation
\begin{eqnarray}
   && \Delta V^*(\eta)V'(\eta) + q \Delta V^*(\eta)  +  p  \Delta V(\eta) =0;\forall \eta \in C ;
\end{eqnarray}

It is equivalent to
\begin{eqnarray}\label{NullVector}
   && C'(\theta) \left(V'(C(\theta))) + q \right)  +  p  C'^* =0;\forall \eta \in C ;
\end{eqnarray}

This equation is simpler than it looks: it is reduced to the total derivative.
\begin{eqnarray}
   \dd{\theta} \left( V(C(\theta))+ q C(\theta) + p C^*(\theta) \right) =0 
\end{eqnarray}

The generic solution is
\begin{eqnarray}\label{Ceq}
    V(C(\theta))+ q C(\theta) + p C^*(\theta) = A
\end{eqnarray}
with some complex constant $A$. 

Plugging it back to the \eqref{Neumann} we have everything cancel except one term
\begin{eqnarray}
   \Im\left(A C'(\theta)\right) =0;\forall \eta =C(\xi)
\end{eqnarray}

The  nontrivial solution for $C(\theta)$ would correspond to $A =0$.

In this case, we arrive at the closed equation for the surface profile $C(\theta)$. This equation is parametric invariant, which allows us to use $\Gamma(\theta)$ as a new parametrization of the loop. 
\begin{eqnarray}
&&\Gamma(\theta) = \gamma x;\\
&& C(\theta) = c(x);\\
&& x \in (0,1);
\end{eqnarray}

The equation reads
\begin{eqnarray}\label{finalCEq}
 q c(y) + p c^*(y) + \frac{\gamma}{2 \pi \i}\Pint_0^1\frac{d x}{c(y)- c(x)}  =0; 
\end{eqnarray}

The principal value prescription here amounts to taking the arithmetic mean of the values of this integral for $\eta = c(y) \pm \i \epsilon c'(y)$. This corresponds to the values inside and outside of the loop $c(.)$.

Although we are considering the 3D problem, our equation almost coincides with the stationary version of so-called \BRE{} \cite{saffman1993} for the 2D interface $c(x)$.

The significant difference is the second term, with $c^*(y)$, making this equation a non-holomorphic function of $c(.)$.

In two dimensions  where the original \BRE{} applies the incompressibility leads to  holomorphic functions.

In our case, the 3D incompressibility leaves the trace of the $x y$  components of the strain arbitrary, so the linear terms can depend upon $\eta^*$.

The real and imaginary parts of a holomorphic function are related by Riemann-Hilbert relations; not so in our case.
Real and imaginary parts of $c(x)$ in our equation are independent functions of $x$.

The original \BRE{} was unstable and did not posses any known fixed points like we are looking for in the \CVS{}.

Our regularization used the exact \BT{} solution of the \NS{} equation in the local tangent plane, and it is stable in 3D provided the \CVS{} conditions.

We can formally reduce it to the unstable solution of conventional \BRE{} in presence of an exponentially growing circulation and an external fluid source, breaking the incompressibility:
\begin{eqnarray}
\d_t Z^*(y,t) = q Z(y,t) + \frac{ \gamma \exp{- 2 p t}}{ 2 \pi \i} \Pint_0^1 \frac{d x}{Z(y,t) - Z(x,t)};
\end{eqnarray}

This equation has an exponentially growing solution related to our \CVS{} fixed point
\begin{eqnarray}
Z(x,t) = c(x) \exp{ - p t};
\end{eqnarray}
Such a solution would not make any physical sense in 2D, but here, in 3D, it is related to the stationary vortex surface.

Let us come back to our \CVS{} stationary equation.

Initially, there were three equations (one for the normal component and another two for the tangent velocity gap annihilating the strain). Our stationary \BRE{} is a complex equation with two unknowns: the real and imaginary parts of the complex curve $c$. 

In our system, one variable $\Gamma(\theta)$ is eliminated by the parametric invariance, and the two remaining unknown variables satisfy the zero strain condition, which follows from \BRE{}. The normal component equation also follows from the \BRE{}.

The function $\Gamma(l)$ in natural parametrization by the length $l$ of the loop is given by the following implicit equation
\begin{eqnarray}
   l = \int_0^{\frac{\Gamma}{\gamma}} \abs{ c'(x)} d x
\end{eqnarray}

The parameter $\gamma$ has the meaning of the period of $\Gamma$, which is the same as conserved velocity circulation $\Gamma$ around the cylindrical tube.
\begin{eqnarray}
   \gamma= \Delta \Gamma = \Re \oint \Delta V(C(\theta)) C'(\theta) d \theta = \oint_C \Delta \vec v(\vec r) \cdot d \vec r
\end{eqnarray}

We need a closed loop, therefore we are looking for a periodic solution of this equation with
\begin{eqnarray}
   c(1) = c(0)
\end{eqnarray}

We could have arrived at equation \eqref{finalCEq} directly by noticing that for a cylindrical geometry, the normal velocity does not have the $z$ component.  

The CVS equation \eqref{finalCEq} simply states that the principal value of the complex velocity vanishes at the surface.  One of the components of this complex velocity is normal to the surface; this component must vanish for the stationary vortex sheet.  Another part is the tangent component of the principal value (arithmetic mean of tangent velocities inside and outside).

By making both components of the principal value of the normal velocity vanish at the surface, we satisfy both  \\CVS{}{} equations.
The tangent strain remains only along the $z$ direction.

The same argument allows us to compute the normal strain on the surface.

The normal vector to the surface is dual to $\Delta \vec v$, the third eigenvector is directed along the $z$ axis.
As one tangent eigenvalue of $S_{\alpha\beta}$ vanishes, its normal eigenvalue equals to minus the remaining tangent eigenvalue $S_{z z}$
\begin{eqnarray}
   S_{n n} = -S_{z z} =  2 p 
\end{eqnarray}

Therefore, independently of the shape of the vortex surface, we have our negative normal strain provided $ p < 0$.

Just as we conjectured \cite{M20c}, it is constant over the surface.

\subsection{Surface equation in detail}

Let us take a closer look at our surface equation. Although it is similar to the stationary \BRE{}, it is not a holomorphic function, which makes it totally different.

By virtue of the periodicity of $c(x)$ we have an identity (for $\eta \in c$)
\begin{eqnarray}
   \frac{1}{2 \pi \i} \Pint_0^1 \frac{d x c'(x)}{\eta - c(x)} = \frac{1}{2 \pi \i}\Pint_c \frac{d \xi }{\eta - \xi} = - \oh 
\end{eqnarray}

Using this identity, we can rewrite our equation in a nonsingular form
\begin{eqnarray}\label{regularCEq}
    &&\mbox{Eq}([c],x)\equiv \nonumber\\
   &&c'(x)\left(q c(x) + p c^*(x)\right) - \frac{\gamma}{2} + \frac{\gamma}{2 \pi\i  }\Pint_0^1\frac{d x' \left(c'(x) - c'(x')\right)}{c(x)- c(x')}  ;
\end{eqnarray}

The Fourier expansion is appropriate for this complex function $c(x)$
\begin{eqnarray}
   c(x) =  \sum_{n =-\infty}^\infty c_n \exp{2 \pi \i n x}
\end{eqnarray}

Note that we do not impose any symmetry requirements on the coefficients like $c_{-n} = c_n^*$, nor $c_{-n} =0$.
There is no such reflection symmetry in our problem.

The Fourier expansion for the complex function $c(x)$ can be viewed as a decomposition into two holomorphic functions
\begin{eqnarray}
  &&c(x) = c_0 + c_+(z) + c_-(1/z);\\
  && z = \exp{2 \pi \i x};\\
  && c_+(z) = \sum_{n=1}^\infty c_n z^n;\\
  && c_-(w) = \sum_{n=1}^\infty c_{-n} w^n.
\end{eqnarray}

Within the numerical approximation, we truncate the Fourier expansion, which is equivalent to the truncation of both of these Taylor expansions for $c_\pm$.

The integral in \eqref{regularCEq}  can be reduced to the contour integral
\begin{eqnarray}
  &&\frac{1}{2 \pi \i}\Pint_0^1\frac{d x}{c(x_0)- c(x)} =\nonumber\\
  &&-\frac{ 1}{2 \pi} \oint \frac{d z}{ z} \frac{1}{c_+(z_0) + c_-(1/z_0) - c_+(z) - c_-(1/z)}
\end{eqnarray}

For any finite number $2 N + 1$ of Fourier coefficients $c_n$ this integral reduces to the sum of residues in the roots of the denominator
\begin{eqnarray}
&&\frac{1}{2 \pi \i}\Pint_0^1\frac{d x}{c(x_0)- c(x)} =\i  \sum_{z \in \mathcal R}  \frac{\theta(1-|z|)}{ (c_+'(z)z - z^{-1}c_-'(z^{-1})};\\
  && \mathcal R: c_+(z_0) + c_-(1/z_0) = c_+(z) + c_-(1/z)
\end{eqnarray}

The roots at the circle, including the root  at $ z = z_0$ enter with $\theta(0) = \oh$. This corresponds to the principal value prescription.

One can rewrite this equation in a form suitable for iteration
\begin{eqnarray}
\label{ck}
  && c_k = \gamma \int_0^1 d x \exp{- 2 \pi \i k x} \left(\frac{ \Re R(x)}{p+ q} + \i \frac{\Im R(x) }{q- p} \right);\\
  \label{Rx}
  && R(x) = \frac{1}{c'(x)} \left(\oh - \frac{1}{2 \pi\i  }\Pint_0^1\frac{d y \left(c'(x) - c'(y)\right)}{c(x)- c(y)}\right)
\end{eqnarray}

This equation possesses a scale invariance
\begin{eqnarray}
&&c(x) \Ra \rho c(x);\\
&& \gamma \Ra \rho^2 \gamma
\end{eqnarray}
which allows us to normalize the solution and trade this normalization for the new dimensionless parameter $\gamma$ to be determined from the equation. We use the standard metric in Fourier space
\begin{eqnarray}
&&  \VEV{a,b } = \sum_k \Re a_k^* b_k =  \int_0^{1} d x \Re a^*(x) b(x);\\
&& \norm c = \sqrt{\VEV{c,c}} = 1
\end{eqnarray}

In the limit of $N\ra \infty$ the norm $\norm a$ becomes the $L^2$ norm  in functional space of complex periodic functions.
We assume smooth closed curves with convergent $L^2$ norm.

Geometrically, for finite $N$, the unit vector $\{c_k\}$ describes a point on a sphere $S_{2 N}$, and the unit vector $v = \{r_k/\norm r \}$ represents the map $\mathcal V: S_{2 N}\mapsto S_{2 N}$. 

We need to find a fixed point $c_\infty$ of this map up to a sign $\sigma$. After that, $\gamma$ is related to the norm of $r[c_\infty]$.
\begin{eqnarray}
&&v[c] =  \frac{r[c]}{\norm {r[c]}} ;\\
&& v[c_\infty] = \sigma c_\infty;\; \sigma = \pm 1\\
&& \gamma_\infty = \frac{\sigma}{\norm{r[c_\infty]}}
\end{eqnarray}

\textit{The Brouwer Theorem  \cite{BrT} would prove the existence of such a fixed point.}

The condition for this theorem is continuity of the map. For any smooth closed curve $c(x)$ its tangent vector $c'(x)$ does not vanish, therefore the function $R(x)$ in \eqref{Rx} remains finite. Should it vanish linearly at some critical point $x_0$, the
Fourier integral in \eqref{ck} would still exist as a principal value, so that the map would remain continuous.

Naturally, we assume that $-q < p < 0$ so that both coefficients  $\frac{1}{q \pm p}$ in our map are finite.

We are leaving to the professionals the rigorous proof and the complete set of assumptions needed to apply the Brouwer theorem.

We take a variational approach: minimize the norm of the equation with an extra condition for the  norm of the solution. The parameter $\gamma$ becomes a variational parameter as well in that case.

\begin{eqnarray}
&&\min_{\{c_k\},\gamma,\lambda} \VEV{c - \gamma r,c - \gamma r } + \lambda \left(\VEV{c,c} -1\right);\\
&& r_k = \int_0^{1} d x \exp{- 2 \pi\i k x}\left(\frac{ \Re R(\theta)}{p+ q} + \i \frac{\Im R(\theta) }{q- p} \right);
\end{eqnarray}

The minimization over $\gamma$ can be performed analytically. 
\begin{eqnarray}
\gamma = \frac{\VEV{c, r}}{\VEV{r, r}}
\end{eqnarray}

The numerical optimization of the vector of Fourier coefficients proved to be too complex to handle by \Mathematica.

In the next publication, we are going to address this problem by means of a large-scale numerical optimization.

\section*{Acknowledgments}

I am grateful to  Sasha Polyakov, Karim Shariff, Katepalli Sreenivasan, and Pavel Wiegmann for inspiring discussions and useful suggestions and to Arthur Migdal for his help with \Mathematica{}.
When I reported the early version of this work on several consecutive seminars by Dennis Sullivan, there was a deep discussion, which revealed a significant bug. I fixed this bug (missing requirement  of vanishing normal velocity in my equations), and I am very grateful to Dennis for the inquisitive spirit of his seminar, reminding me of Landau and Gelfand seminars.

This work is supported by a Simons Foundation award ID $686282$ at NYU. 
 \bibliography{bibliography}

\end{document}